\documentclass[epjST]{svjour}

\usepackage{graphicx}
\usepackage{amsmath}
\usepackage[utf8]{inputenc}
\usepackage{amsmath}
\usepackage{amsfonts}
\usepackage{amssymb}
\usepackage{graphicx}
\usepackage{forloop}

\usepackage{color}   
\usepackage{float}

\usepackage{multirow}
\usepackage{multirow}
\usepackage{amsmath}
\usepackage{amsfonts}
\usepackage{amssymb}
\usepackage{graphicx}
\usepackage[section]{placeins}
\begin{document}
\title{Phase Separation of Superfluids in the Chain of Four-Component Ultracold Atoms}
\author{G. Barcza\inst{1}\fnmsep\thanks{\email{barcza.gergely@wigner.mta.hu}} \and E. Szirmai\inst{2}\fnmsep\thanks{\email{eszirmai@phy.bme.hu}}
 \and J. S\'olyom\inst{1} \and \"O. Legeza\inst{1}}
\institute{Strongly Correlated Systems "Lend\"ulet" Research Group, Wigner Research Centre for Physics, HAS,  H-1525 Budapest, Hungary 
\and BME-MTA Exotic Quantum Phases "Lend\"ulet" Research Group, Budapest University of Technology and Economics, Institute of Physics, H-1111 Budapest, Hungary}
\abstract{
We investigate the competition of various exotic superfluid states in a chain of spin-polarized ultracold fermionic atoms with hyperfine spin $F = 3/2$ and s-wave contact interactions.
We show that the ground state is an exotic inhomogeneous mixture in which two distinct superfluid phases  --- spin-carrying pairs and  singlet quartets --- form alternating domains in an extended region of the parameter space. 
} 
\maketitle
\section{Introduction}
\label{intro}
Superfluidity is an experimentally well studied inherent feature of ultracold atomic systems \cite{bloch08a}.
Recent experiments  \cite{liao10a} have demonstrated that the normal state competes with superfluid pairing states in one-dimensional two-component homonuclear Fermi gases for finite population imbalance of the different spin states exhibiting the Fulde-Ferrell-Larkin-Ovchinnikov (FFLO) state \cite{fulde64a}.
Investigation of heteronuclear binary mixtures have also revealed novel superfluid states \cite{liu3a} and exotic phases \cite{iskin,baarsma}. 
Effective Hamiltonians with local contact interactions have proven to be instrumental in understanding the physics of such systems. 
Spin-imbalanced two-component Hubbard model with attractive coupling has been applied to study pairing in homonuclear binary mixtures \cite{fflo} while asymmetric Hubbard model has been used to describe heteronuclear systems \cite{barberio10a,dalmonte12a}. 
The Falicov-Kimball model \cite{falicov} can be considered as a special case of the asymmetric Hubbard model where the  heavy particles are immobile and represent an external potential for the itinerant particles \cite{ates05a,maska11a}. 

The four-component (spin-3/2) Fermi gas with contact interactions can provide even more exotic superfluid states.
Their phase diagram  is well established in case of spin balance~\cite{wu06a,capponi07a}.
Assuming local interactions, Pauli's exclusion principle permits only singlet ($S=0$) and quintet ($S=2$) collisions. 
In this case the Hamiltonian of a system with $L$ sites can be written as
\begin{equation}
{\cal H} = -t\sum_{i=1,\alpha}^L \big( \hat{c}_{\alpha,i}^\dagger \hat{c}_{\alpha,i+1}^{\phantom\dagger} + {\rm H.c. }\big) + 
g_0 \sum_{i=1}^L \hat{P}_{0,i} + g_2 \sum_{i=1}^L \hat{P}_{2,i}\,.
 \label{eq:ham1}
\end{equation}
Here $t$ measures the one-particle overlap between neighboring sites, $g_S$ denotes the coupling parameter in the spin-$S$ channel and the operator $\hat{P}_{S,i}$ projects to the corresponding subspace at site $i$.
$\hat{P}_{S,i}$ can be expressed as 
$\hat{P}_{S,i}=\sum_{m} \hat{P}_{Sm,i}^\dagger \hat{P}_{Sm,i}^{\phantom\dagger}$,
where the creation operator of a pair with total spin $S$ and $z$ component $m$ on site $i$, $\hat{P}_{Sm,i}^\dagger$, can be given in terms of the Clebsch-Gordan coefficients with the standard notation as
$\hat{P}_{Sm,i}^\dagger = \sum_{\alpha,\beta} \left< \frac32, \frac32; \alpha, \beta |S,m\right> 
\hat{c}_{\alpha,i}^\dagger \hat{c}_{\beta,i}^\dagger$.

For strong enough attractive quintet coupling ($g_2<0$), the ground state of the quarter-filled spin-balanced system is described by site-centered spin-singlet quartets in an extended regime of the singlet coupling  $g_0$~\cite{wu06a,capponi07a}. 
In spin-polarized system, for finite $\tilde{m}=1/L\sum_{i,\alpha} \alpha n_{\alpha,i}$, the spin-singlet quartets are expected to be replaced at least partially by other energetically favorable spin-carrying excitations.
We have shown in Ref. \cite{barcza12a} that the total number of quintet pairs with $m=2$ increases as a function of $\tilde{m}$ while the number of quartets decreases progressively and  disappears completely for $\tilde{m}=1$ in the $0<g_0<-3g_2$, $g_2<0$ regime. 
We have observed that the correlation functions of both the quartets and quintet pairs decay algebraically for small or moderate spin imbalance, their superfluid quasi-condensates may coexist by displaying quasi-long range ordering in this regime.

In this paper, we extend our previous work \cite{barcza12a} by investigating the ground state properties of the system as a function of spin polarization for such couplings.
We shall point out that the ground state is in fact a mixture of the two superfluid phases and shall analyze various features of this mixture by means of density profiles, correlation functions and momentum distributions. 
In the following, we denote correlation of the $m=2$ pairs as $\chi_{P}(i,i') = \langle P_{22,i}^{\dagger}P_{22,i'}^{\phantom\dagger} \rangle$. 
Correlation of the spin-singlet quartets is $\chi_{Q}(i,i') = \langle \hat{Q}_{i}^{\dagger} \hat{Q}_{i'}^{\phantom\dagger}
\rangle$ with $\hat{Q}_{i}^{\dagger}=\hat{c}^{\dagger}_{3/2,i}\hat{c}^{\dagger}_{1/2,i}\hat{c}^{\dagger}_{-1/2,i}\hat{c}^{\dagger}_{-3/2,i}$.
The local density of $m=2$ pairs is measured by the exclusive pair densities $\hat{N}_{P,i}$ expressed in terms of the pairing operators , e.g., $\hat{N}_{P,i}=\prod_{\alpha<0}(1-\hat{n}_{\alpha,i}) \prod_{\alpha'>0}\hat{n}_{\alpha',i}$, while the local quartet density operator reads $\hat{N}_{Q,i}=\hat{n}_{-3/2,i}\hat{n}_{-1/2,i}\hat{n}_{1/2,i}\hat{n}_{3/2,i}$.

We applied  the density matrix renormalization group (DMRG)  method \cite{dmrg} in order  to investigate systems up to $L=64$ sites keeping $500-2000$ block states and using 10 sweeps.
The truncation error was usually smaller than $10^{-7}$.

\section{Phase separated domain structure}

\begin{figure}[!tb]
\centering
 \hskip -0.35cm
    \includegraphics[height=32.5mm,width=44.5mm]{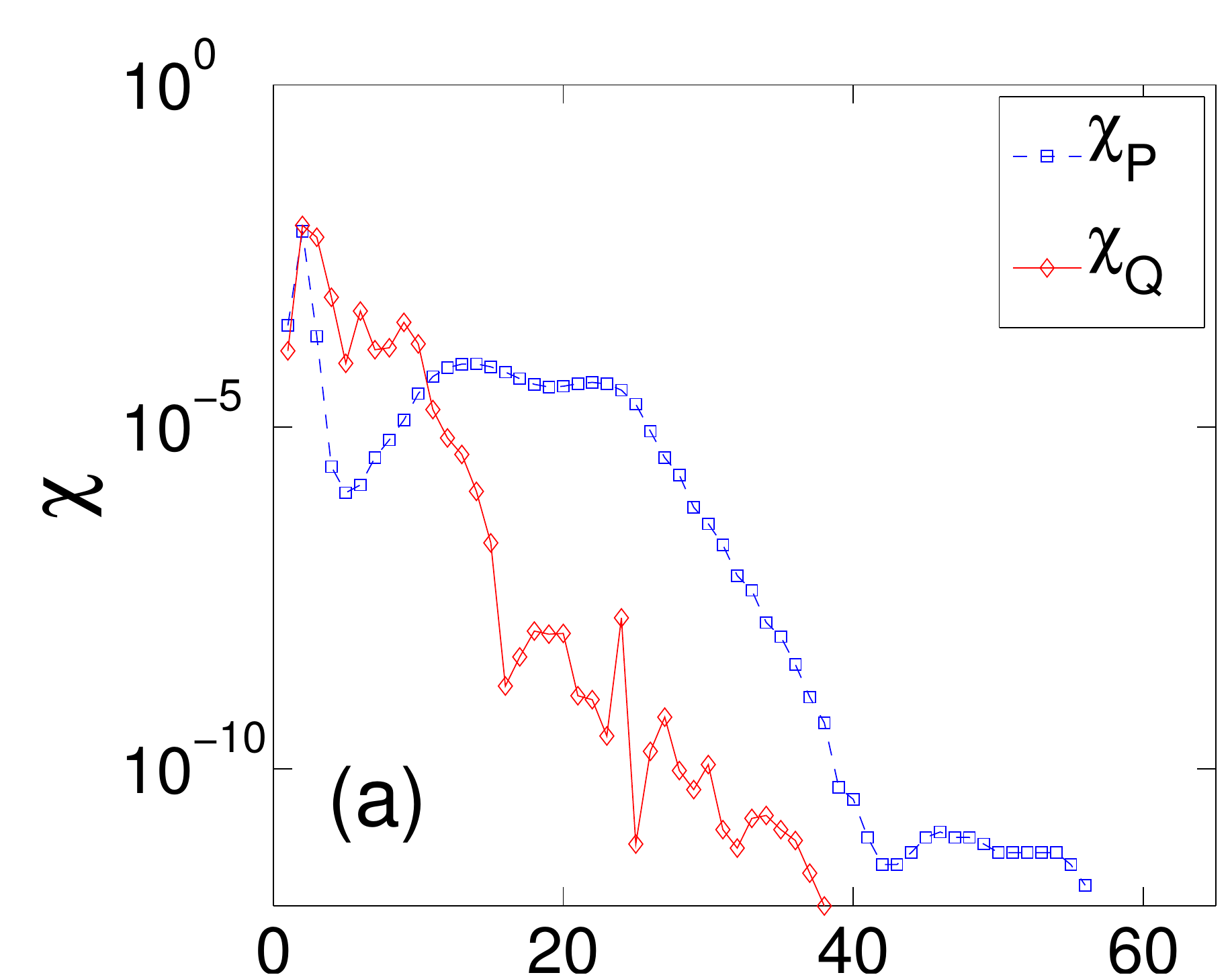}
    \hfill
    \includegraphics[scale=0.22]{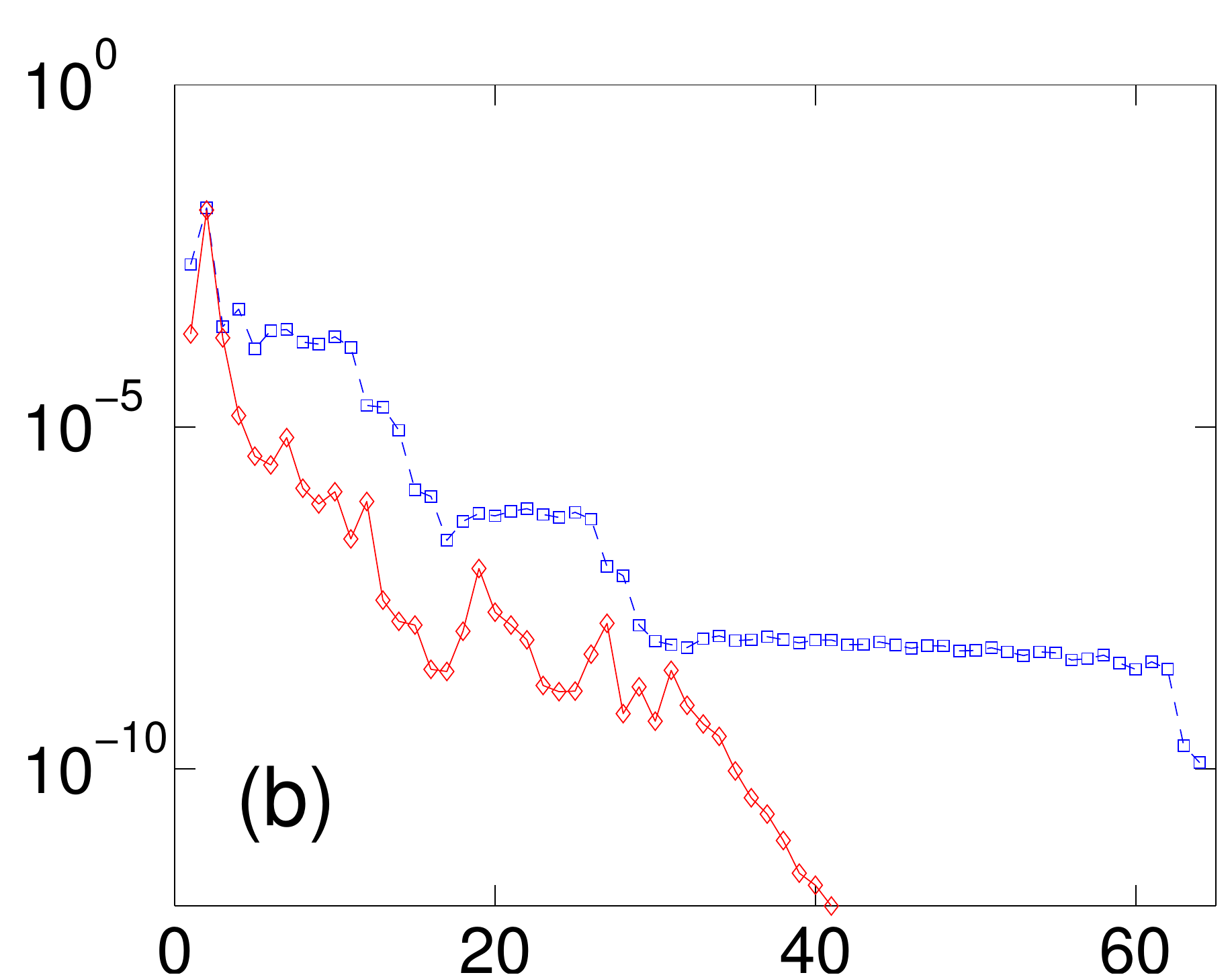}
  \hfill
  \includegraphics[scale=0.22]{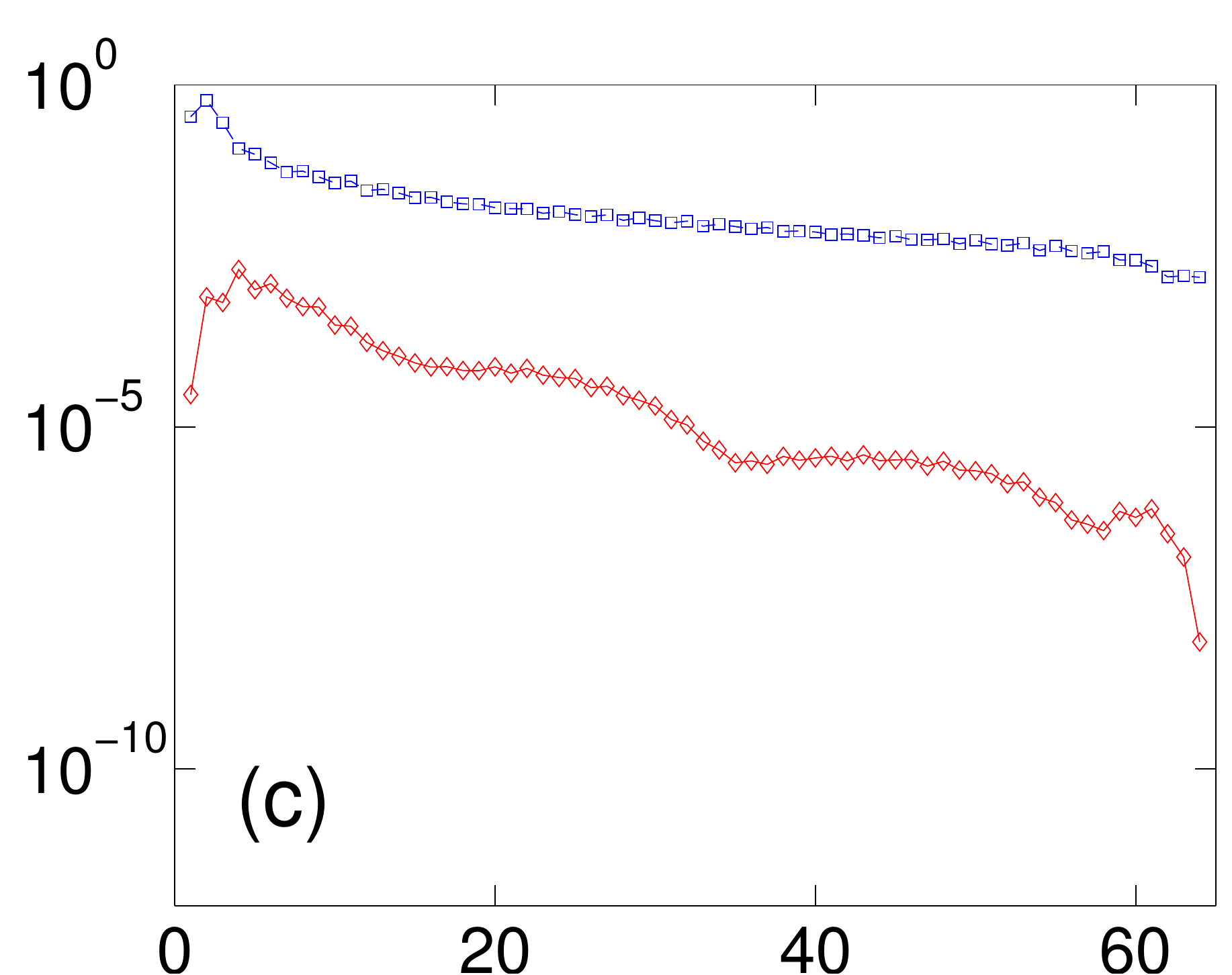}\\
    \includegraphics[scale=0.22]{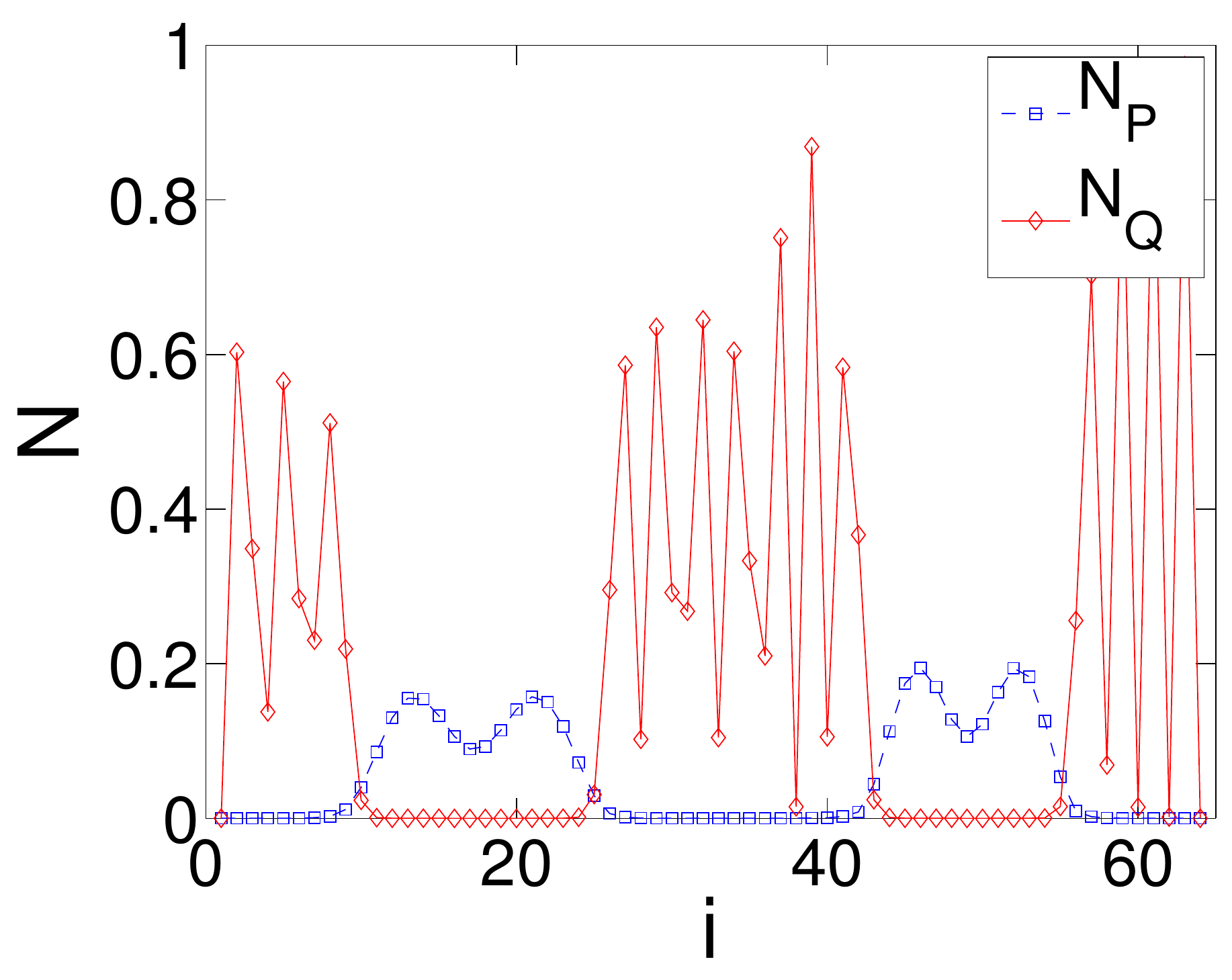}
  \hfill
  \includegraphics[scale=0.22]{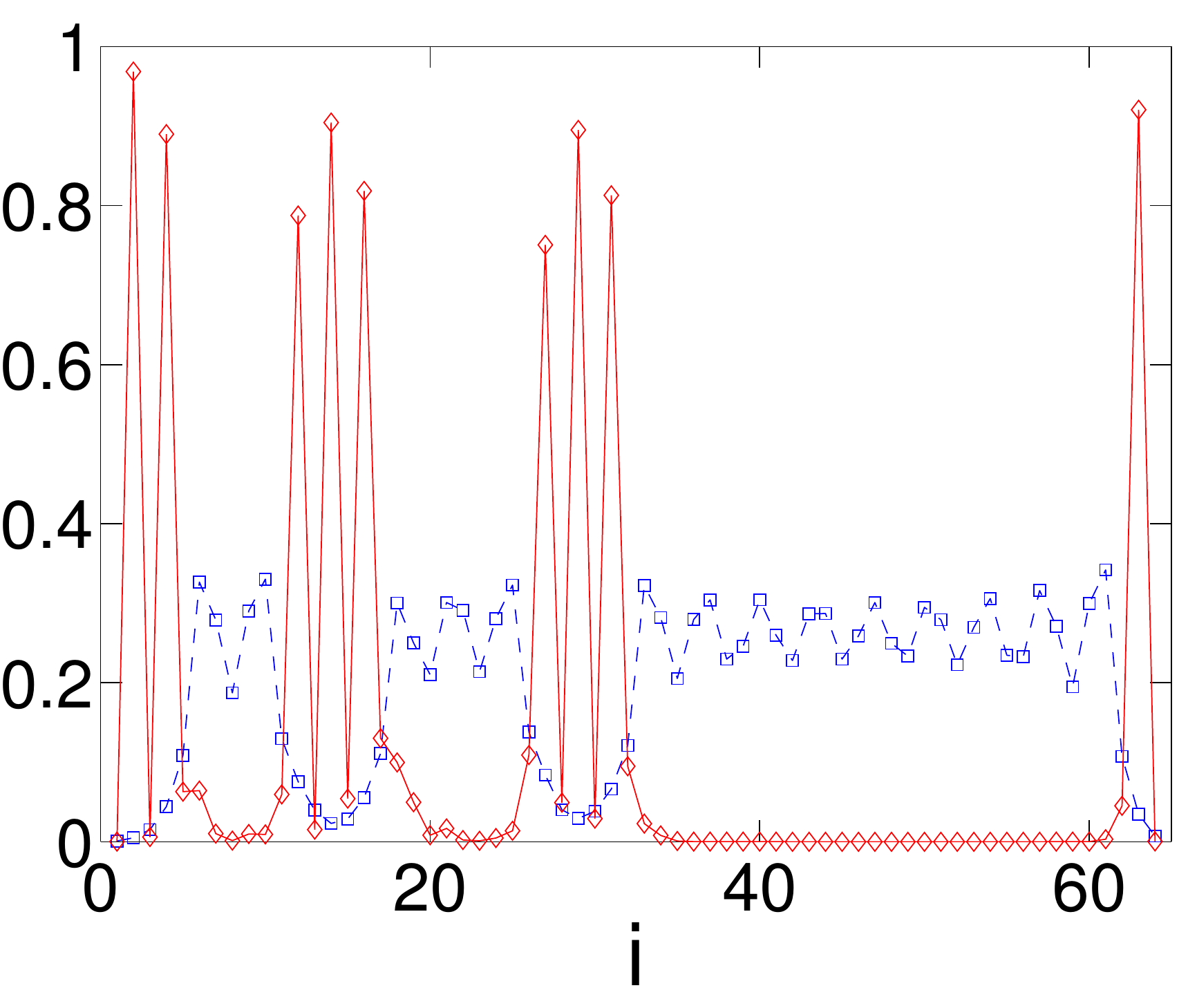}
  \hfill
  \includegraphics[scale=0.22]{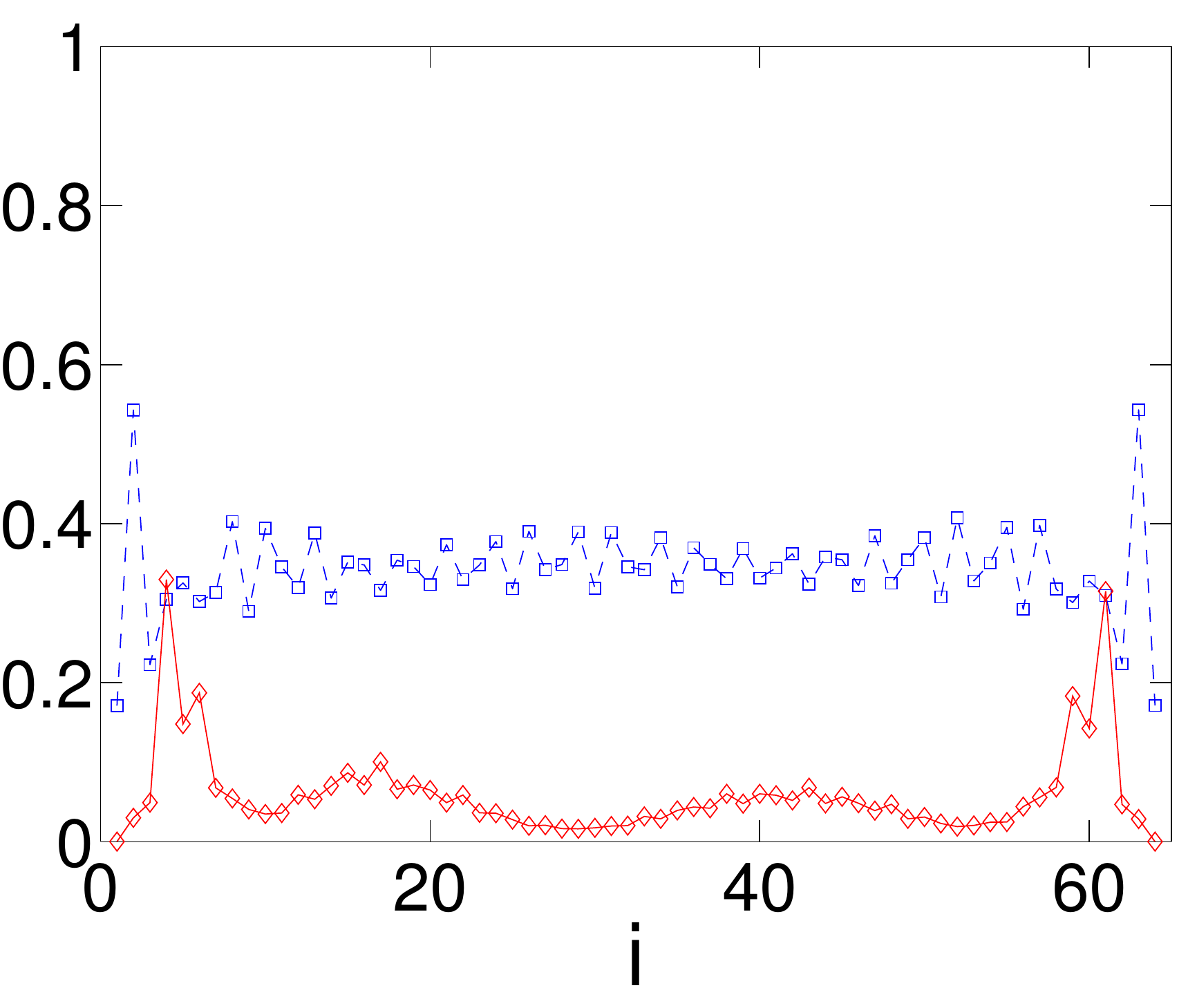}  
  \vskip -6.5cm
  \hskip 2.2cm
   \includegraphics[scale=0.11]{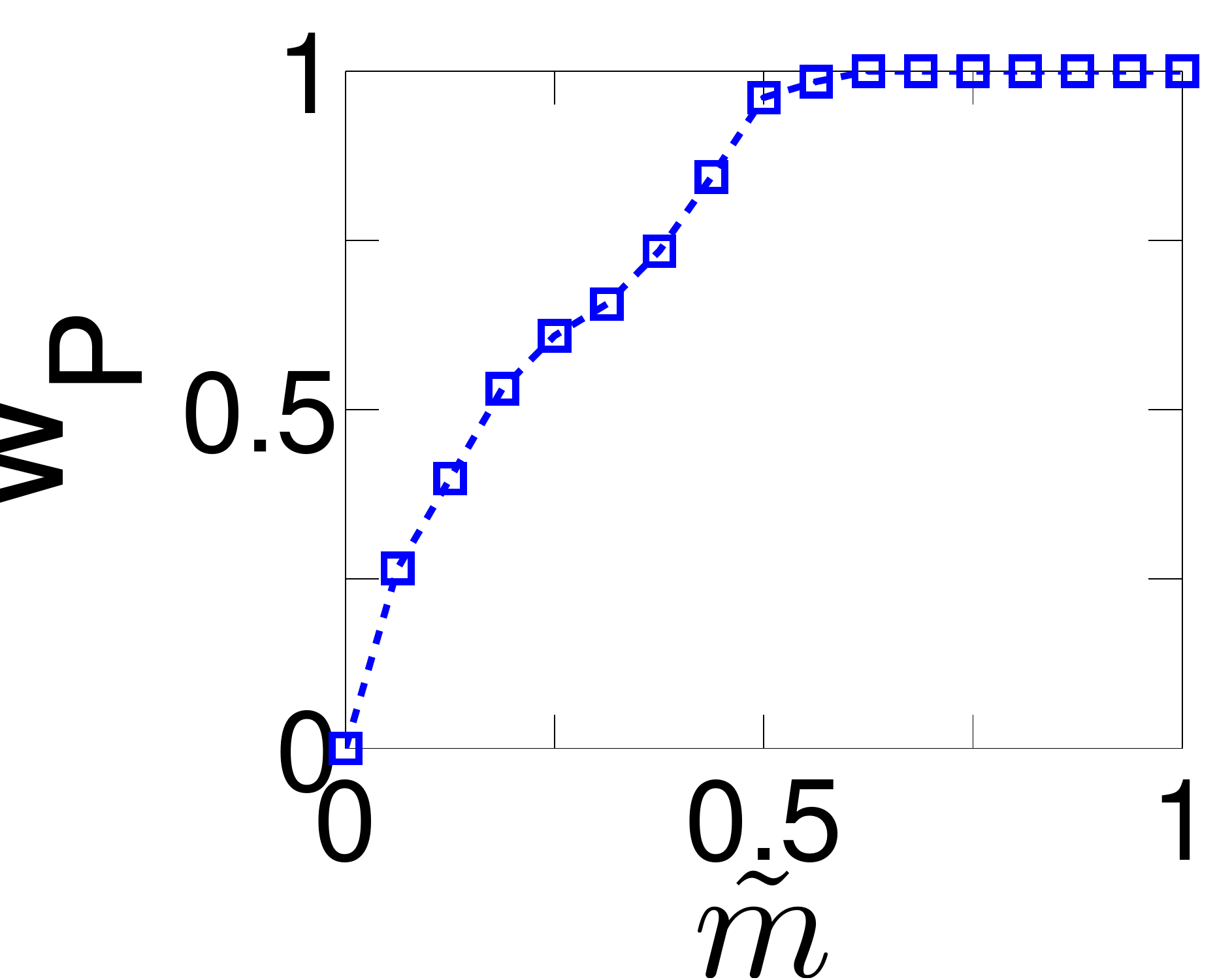} 
  \vskip 4.9cm
\caption{(Color online) Correlation functions and spatial density profile of $m=2$ quintet pairs and quartets measured on chain of $L=64$ sites at (a) $\tilde{m}=1/8$, (b) $\tilde{m}=7/16$, (c) $\tilde{m}=3/4$ $(g_0=2, ~ g_2=-4)$.
The inset depicts the overall width of the $m=2$ pair condensate $w_P$, in units of the chain length, as a function of spin polarization.}
\label{fig:density_profile_g0_2g2_-4}
\end{figure}

Investigating the $m=2$ pair-pair and quartet-quartet correlation functions depicted in Fig.~\ref{fig:density_profile_g0_2g2_-4} as function of magnetization, we observe that for weak polarizations both correlations show a staggered, step-like structure superimposed on the algebraic decay.
This inner structure can be better revealed by studying the corresponding density profiles plotted also in Fig.~\ref{fig:density_profile_g0_2g2_-4}.
The spatial distribution of quartets and quintet pairs clearly shows that these superfluid condensates are segregated.
As a result, the correlation function decays algebraically in the regions filled by the associated condensate while it drops rapidly in the regions where the complementer phase dominates.

Tuning the actual coupling and spin polarization, the number and average length of the domains varies drastically due to the delicate balance of the domain-wall energy and the kinetic energy which increases with the size of the domain.
Such segregation of phases is not exceptional in low-dimensional systems with two components: in models with different mobilities --- such as the Falicov-Kimball model \cite{ates05a,maska11a,cencarikova12a} or the strongly asymmetric Hubbard model \cite{barberio10a} --- the ground state can be segregated exhibiting irregular arrangements of domains. 
The spin-carrying pairs with high mobility exert a pressure on the domains of the quartets. 
This pressure can also be interpreted as an external confining potential acting on the quartets condensed into the domains. 
The behaviour of a one-dimensional system consisting only of quartets was studied in Ref.\cite{roux} in presence of external potential. 
It was found that for a weak confining potential the quartets are delocalized. 
For stronger confinement the quartets are condensed into a finite region of the chain with alternating empty and occupied sites. 
The analogous behaviour of the quartets can be observed in Fig.~\ref{fig:density_profile_g0_2g2_-4} for $\tilde{m}=1/8$ and $7/16$, respectively, as the increasing spin polarization (and accordingly the number of the pairs) leads to larger effective pressure. 
Further increasing $\tilde{m}$, the total number and energy contribution of the quartets is reduced and the width of their condensate progressively shrinks in favor of the expanding quintet pairs as presented in the inset of Fig. \ref{fig:density_profile_g0_2g2_-4}.
For large enough spin polarization, above a critical value $\tilde{m}_c$, the segregation of the phases disappears and both superfluid condensates are smeared over the whole lattice while the corresponding correlation functions decay smoothly, see column (c) of Fig.~\ref{fig:density_profile_g0_2g2_-4}.

While the coexistence of the quartets and the spin-carrying pairs characterizes the system in the whole region $0<g_0<-3g_2$, $g_2<0$, this domain structure disappears at around $g_0 \approx -2g_2$ as $g_0$ increases. 
Since the quartets become less robust just as less classical for relatively large $g_0$, they coexist with the quintet pairs without phase separation even for weak spin polarization. 

\section{Critical magnetization}

As was mentioned, the domain structure of the mixed superfluid phase disappears above a critical value $\tilde{m}_c$.
From the numerical analysis, we conclude that $\tilde{m}_c\approx 1/2$ and it does not depend on the couplings $g_0$, and $g_2$ just as it is not sensitive to the length of the chain $L$. 
We emphasize that the domain structure is absent for $\tilde{m}_c \leq \tilde{m} <1$, even though quartets and $m=2$ pairs are present simultaneously and their correlation functions exhibit identical smooth algebraic decay.
Here we present further evidences of such a transition in experimentally measurable quantities.  

\begin{figure}[!h]
\centering
    \includegraphics[scale=0.25]{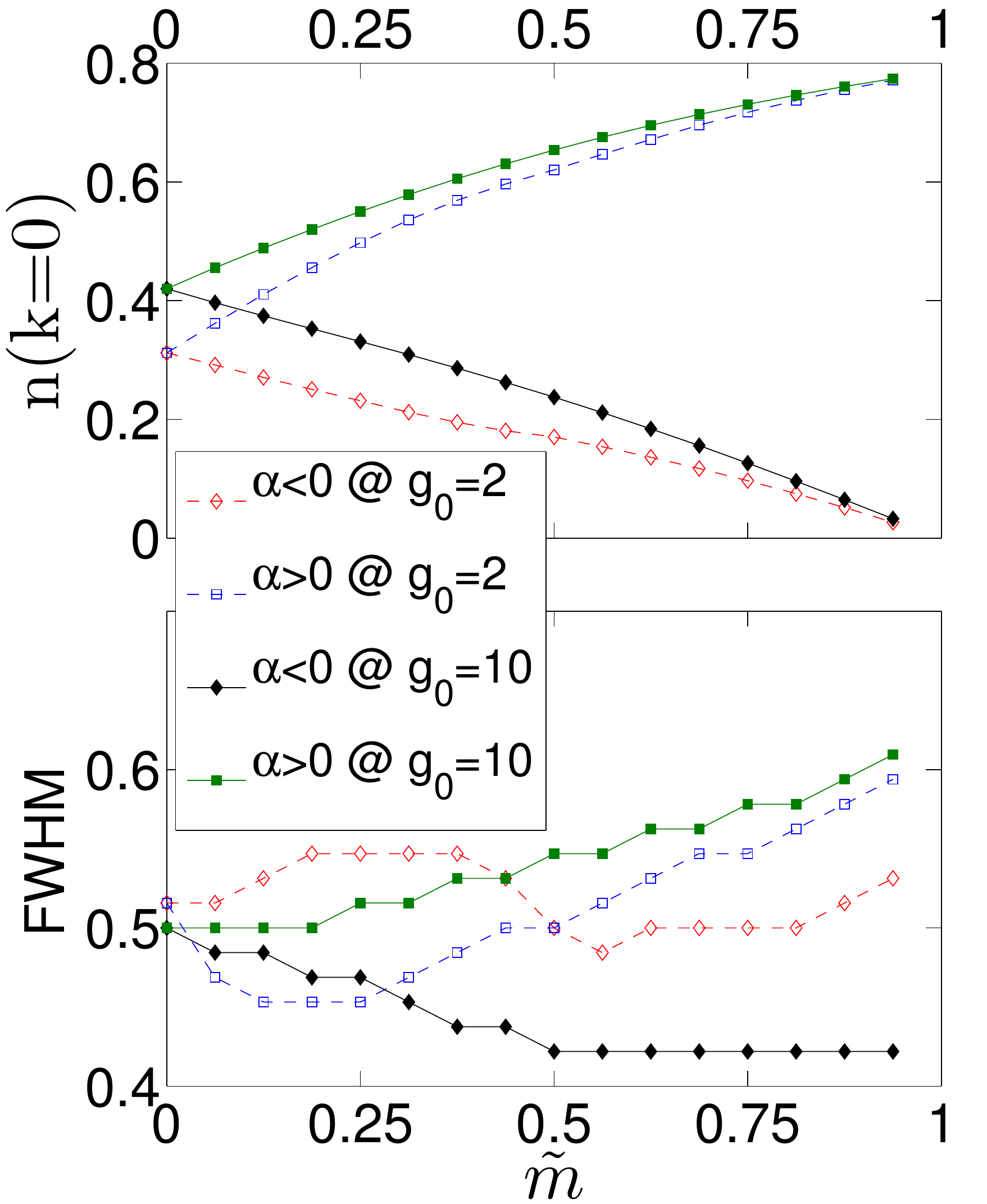}
  \hskip 0.7cm
  \includegraphics[scale=0.25]{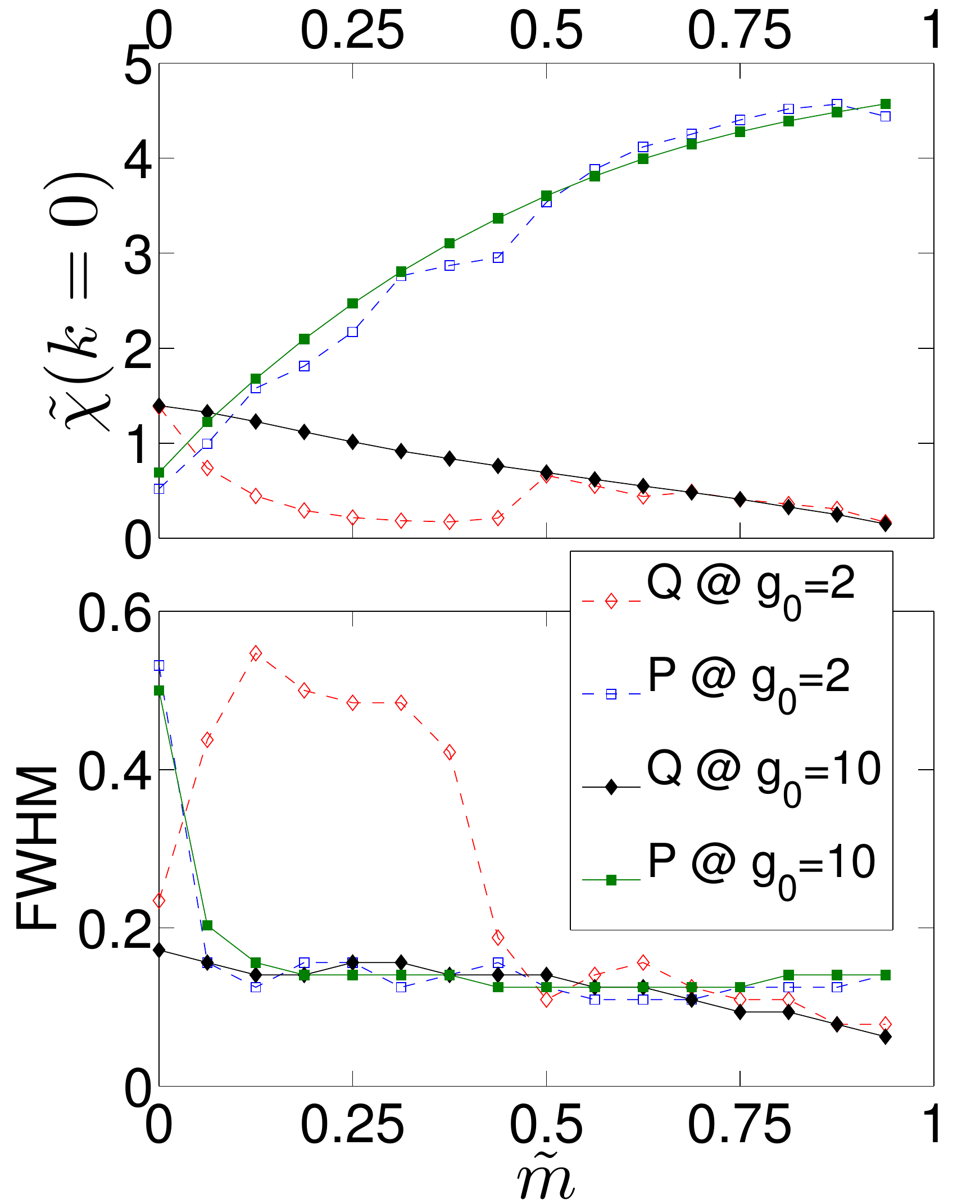}\\
\caption{(Color online) (Left panel) Height of zero momentum polarization peak of  $\tilde{n}_\alpha(k)$ and corresponding FWHMs (measured in units of inverse lattice spacing) in the regime of $0\leq \tilde{m}<1$  for $(g_0=2, ~ g_2=-4)$ and $(g_0=10,~g_2=-4)$. 
(Right panel) Similar quantities for  $\tilde{\chi}_P$ and $\tilde{\chi}_Q$.}
\label{fig:momentum}
\end{figure}

Applying Fourier analysis \cite{dalmonte12a}, first we study the momentum distribution for each spin component of the fermions via the Fourier transform of the one-body density matrix, $\tilde{n}_\alpha(k) = 1/L\sum_{j,j'} e^{ik(j-j')} \langle \hat{c}^\dagger_{\alpha,j} \hat{c}^{\phantom\dagger}_{\alpha,j'} \rangle$. 
We find that the  center of mass (COM) momentum $k_{\rm max}$ of the densities is zero irrespective of $\tilde{m}$.
The $k=0$ height of densities increases identically as a function of $\tilde{m}$ for positive spin projection $\alpha=1/2,~3/2$. 
Results are presented in the left panel of Fig. \ref{fig:momentum} for $g_0=2$ and $g_0=10$ at $g_2=-4$ where phase separation is present or absent, respectively.
Similarly, the peak of the densities for negative spin projection  $\alpha=-3/2,~-1/2$ decreases with identical rate. 
These tendencies are independent of the couplings $g_0$ and $g_2$, and show no signature of any transition. 
The Fourier components are finite even for momenta close to the edge of the Brillouin zone.

In order to quantify the coherence of the condensate, we define the full width at half maximum (FWHM) of the distribution as the distance at which the function reaches half of its image range. 
We find that the coherence does not change drastically as the polarization is increased but it has a characteristic feature: the monotonical variation of the coherence or decoherence revealed by the variation of FWHM breaks down when domain structure is formed.

Even more conspicuous signature of the phase transition can be observed in the momentum distribution of the $m=2$ pair and quartet condensate defined as 
$\tilde{\chi}_{P}(k) = 1/L\sum_{j,j'} e^{ik(j-j')} \chi_{22}(j,j')$,  and
$\tilde{\chi}_Q(k) = 1/L\sum_{j,j'} e^{ik(j-j')} \chi_{Q}(j,j')$,
respectively.

We find that the momentum distribution of both condensates decreases monotonically from the center to the edge of the Brillouin zone for a fixed value of $\tilde{m}$ , and the COM momentum remains zero up to $\tilde{m}=1$. 
The zero COM momentum indicates that these spin-carrying pairs are not FFLO-like. 
Thanks to the commutation relation of the associated superfluid creation operators, $\hat{P}^\dagger, \hat{Q}^\dagger$, and to the constraint on their occupation number, the condensates behave approximatively like hard-core bosons in the low-density limit.
Accordingly, the emerging state can be considered as a repulsing mixture of hard-core boson-like superfluids \cite{hu,guglielmino11a,roscilde} irrespective of the relative weights of the components.

On the contrary, for $\tilde{m} \geq 1$, fermions with negative spin projection are frozen out, therefore the system can be described by the half-filled spin-imbalanced spin-1/2 Hubbard model featuring the finite momentum FFLO phase \cite{fflo}.
In this phase, the COM momentum of pairs with $m=2$ increases linearly with evolving spin polarization. 

In the right panel of Fig. \ref{fig:momentum} we show the spin-polarization dependence of the height of quartet and pair condensate peaks $\tilde{\chi}_Q(k=0)$ and $\tilde{\chi}_P(k=0)$, respectively. 
Increasing the spin-polarization, the weight of the $k=0$ component decreases for the quartets while that for the pairs increases monotonically for weak quartetting (see Fig. \ref{fig:momentum} at $g_0=10$).
In case of phase separation (see Fig. \ref{fig:momentum} at $g_0=2$), the behaviour of the $k=0$ peak of the pair-pair correlation shows similar behaviour as in the previous case. 
The $k=0$ quartet-quartet correlation drops significantly as the spin imbalance starts to increase and develops a jump around $\tilde{m}_c$ providing a well distinguishable signature of the transition. 
As the polarization is further increased, $\tilde{\chi}_Q(k=0)$ decreases just as in case of $g_0=10$. 

Similar unambiguous characteristic feature can be observed in the behaviour of the FWHM of the quartet-quartet correlation function below the critical polarization. 
The FWHM of the $m=2$ pair correlation function is roughly $1/10$ for any finite polarization independent of the couplings. 
In contrast to this, the FWHM of the quartets becomes as large as $1/2$ in the segregated condensate reflecting the crystallization, while above the critical polarization $\tilde{m}_c$ the FWHM of the quartet condensate, too, drops to $1/10$.

\section{Summary}
In this paper, based on our earlier work on spin-polarized four-component interacting one-dimensional Fermi gases~\cite{barcza12a}, we further investigated the properties of the superfluid mixture by studying the characteristic features of the density profiles of the condensates.
We found that in an extended region of the parameter space a phase separated state of spin-carrying two-particle pairs and spin-singlet quartets emerges below a critical value of the spin polarization. 
We showed that the existence of the domain structure can be detected unambiguously in various experimentally measurable quantities.

\section*{Acknowledgments}
This research was supported in part by the Hungarian Research Fund (OTKA) under Grant Nos. K 100908 and NN110360. 
The authors acknowledge computational support from Philipps Universit{\"a}t, Marburg. 
G. B. thanks Á. Bácsi for the fruitful discussions.

\end{document}